# Building Capacity for Artificial Intelligence in Africa: A Cross-Country Survey of Challenges and Governance Pathways

**Aryee**, J. N. A.[1,*], **Davies**, P.[1], **Torsah**, G. A.[1], **Apaw**, M. M.[1], **Boateng**, C. D.[2], **Mwando**, S. M.[3], **Kwisanga**, C.[4], **Jobunga**, E.[5], **Amekudzi**, L. K.[1]

[1] *Department of Meteorology and Climate Science, FPCS, COS, KNUST*
[2] *Department of Physics, FPCS, COS, KNUST*
[3] *Namibia University of Science and Technology, Khomas, Windhoek, Namibia*
[4] *University of Rwanda, College of Science and Technology, Kigali, Rwanda*
[5] *Department of Mathematics and Physics, Technical University of Mombasa, P.O. Box 90420 - 80100, Mombasa, Kenya.*

**ORCID**
**Aryee**, J. N. A.:  0000-0002-4481-1441
**Torsah**, G. A.:  0009-0000-0556-2959
**Davies**, P.:  0000-0001-9598-4652
**Apaw**, M. M.:   0009-0008-7674-8433
**Mwando**, S. M.:   0000-0002-5993-2764
**Kwisanga**, C.:  0000-0002-3541-1542
**Boateng**, C. D.:  0000-0002-1721-4158
**Jobunga**, E.:
**Amekudzi**, L. K.:  0000-0002-2186-3425

Corresponding Author: jnaaryee@knust.edu.gh

**Abstract**
Artificial intelligence (AI) is transforming education and the workforce, but access to AI learning opportunities in Africa remains uneven. With rapid demographic shifts and growing labour market pressures, AI has become a strategic development priority, making the demand for relevant skills more urgent. This study investigates how universities and industries engage in shaping AI education and workforce preparation, drawing on survey responses from five African countries (Ghana, Namibia, Rwanda, Kenya and Zambia). The findings show broad recognition of AI's importance but limited evidence of consistent engagement, practical training, or equitable access to resources. Most respondents who rated the AI component of their curriculum as very relevant reported being well prepared for jobs, but financial barriers, poor infrastructure, and weak communication limit participation, especially among students and underrepresented groups. Respondents highlighted internships, industry partnerships, and targeted support mechanisms as critical enablers, alongside the need for inclusive governance frameworks. The results showed both the growing awareness of AI's potential and the structural gaps that hinder its translation into workforce capacity. Strengthening university-industry collaboration and addressing barriers of access, funding, and policy are central to ensuring that AI contributes to equitable and sustainable development across the continent.

1. Introduction

Artificial intelligence (AI) is reshaping economies and redefining how knowledge is created and applied (**Colther and Doussoulin, 2024**). Across sectors such as agriculture, health, finance, and education, AI promises new solutions to persistent developmental challenges (**Azaroual, 2025**; **Aijaz et al., 2025**). Africa's fast-growing youth population and persistent gaps in employment, productivity, and governance make the adoption of AI more than a technological choice, but a strategic issue for development (**World Bank report, 2014**; **Azaroual, 2025**). Yet Africa's ability to benefit from AI depends on building local expertise and aligning education systems with labour market needs. Higher Education Institutions (HEIs) and industries hold central roles in this task, and their collaboration will determine how well the workforce can meet the demands of an AI-driven economy.

The interface between universities and industry has been a longstanding concern in African higher education, but the rise of digital technologies has amplified its urgency. Many academic programmes still emphasise theory with little exposure to applied industry challenges, producing graduates who struggle to meet employer expectations (**Odetola et al., 2018**). In AI, the gap is wider , given the rapid pace of change and reliance on computational tools, datasets, and real-world experimentation (**Jembu and Lee, 2025**). Without strong university-industry linkages, AI education risks becoming detached from the skills required in professional settings or too fragmented to build lasting capacity.

Recent initiatives such as AI4D Africa, innovation hubs, and sector-specific pilots show growing interest in AI for development. However, these efforts often face challenges of scale, sustainability, and equity. Universities continue to experience shortages in computing infrastructure, specialised faculty, and funding for practical education (**Aydin, 2021**). Meanwhile, as industries begin to adopt AI, they lack structured mechanisms to share their skill demands with academic institutions (**Mastercard Report, 2025**). This disconnect undermines both workforce preparation and the capacity of HEIs to align curricula with emerging needs.

Understanding how perceptions, engagement, and access to resources influence AI education is essential for strengthening these linkages. Perceptions shape how students and staff value AI's potential and motivate their investment in related skills. Engagement captures the partnership between universities and industry in creating opportunities for students to translate theory into practice through coursework, research, internships, and collaborative projects. Access to resources such as computing facilities, datasets, and online materials determines whether students or workers can meaningfully participate in AI activities. Together, these elements define the environment in which universities prepare future AI professionals and respond to industry expectations.

In Africa, disparities in access to AI education and resources remain stark. Students at urban universities benefit from stronger availability of AI-related courses and computing infrastructure, while those in smaller or rural institutions face limited opportunities. These inequities risk widening skills gaps and undermining inclusive workforce development, underscoring the need for targeted investments in infrastructure, faculty capacity, and industry–university partnerships (**Li, 2024**; **Mokoena & Seeletse, 2025**). Faculty shortages and unstable funding constrain the ability of many programmes to sustain applied projects or partnerships with industry (**Jonbekova et al., 2020**). At the same time, industries report difficulty finding graduates with practical AI skills, despite evidence of strong interest among students (**Ramos et al., 2025**). This suggests that awareness of AI's relevance is growing, but structured pathways to translate that awareness into expertise are still weak.

Empirical research on these dynamics in African contexts is limited. Most studies on AI education and workforce preparation draw on experiences from high-income regions, where infrastructure is stronger and industry ecosystems are mature (**Cross & Feldman, 2025**). Africa's context is distinct, with uneven AI adoption, fragmented industrial bases, and higher education systems striving to overcome legacy capacity gaps while keeping pace with rapidly evolving technologies (**Maluleke, 2025**; **Sangwa et al., 2025**). Region-specific studies are therefore needed to examine how perceptions, practices, and resource availability shape AI education and how collaboration between HEIs and industry can accelerate progress. This study responds to this need by assessing university-industry linkages for AI education in Africa through the lens of perceptions, engagement, participation, and access to learning resources. Drawing on survey data, the analysis explores respondents' perspectives on AI's future impact, their participation in related activities, and the resources available to support their learning. These findings are situated within the broader goal of strengthening the complementary roles of universities and industries in developing a workforce equipped to advance AI in ways that reflect Africa's strategic priorities. They point to an encouraging recognition of AI's importance, while also revealing persistent gaps in practice and infrastructure that require coordinated action among universities, industries, and policymakers.

## 2. Methodology
### 2.1. Research Design

The study adopted a mixed-methods approach to explore how university curricula across Africa align with the evolving demands of AI-related workforce development. The study combined quantitative survey analysis with qualitative thematic mapping to capture a comprehensive picture of institutional engagement, curriculum relevance, and perceived skill gaps. The survey questionnaire was developed through an iterative process informed by existing literature on AI education, equity frameworks, and peer feedback. It was structured around five strategic research questions: (i) how well curricula align with workforce needs, (ii) how universities and industries collaborate on AI skills development, (iii) what factors affect equity and access in AI education, (iv) how perceptions and resources shape engagement, and (v) what policy actions can advance inclusive AI education across Africa. Each forms a distinct module within the survey, allowing for targeted analysis and thematic synthesis. The survey included a blend of multiple-choice, Likert-scale, and open-ended questions, which ensured cultural and contextual relevance. These were tailored to different respondent roles, such as faculty, curriculum designers, and administrators and incorporated branching logic to guide participants through the most relevant sections. The questionnaire was reviewed and approved by the institutional ethics board (Committee on Human Research, Publication and Ethics; **ref: CHRPE/AP/525/25**) in accordance with established research ethics standards, ensuring compliance prior to its administration.

## 2.2. Sampling and data collection

Data collection was carried out online for about 2 months, with outreach facilitated through university networks. A purposive sampling strategy was used to ensure representation across regions and institutional types, including public and private universities, technical institutes, and emerging AI hubs. The final dataset included responses from a diverse range of stakeholders, offering rich insights into both structural and experiential dimensions of AI education.

## 2.3. Analysis

The survey data contained multiple Likert-type scales that varied across questions. To enable consistent analysis and aggregation, all categorical responses were harmonised and encoded into a common ordinal scale ranging from 0 to 4. The harmonisation involved two key steps. Categorical options such as frequency-based responses were re-ordered according to increasing intensity. For example, the categories Never, Occasionally, Regularly, Frequently, and Always were mapped to the numerical scale 0-4. Similarly, responses such as Strongly Disagree, Disagree, Neutral, Agree, Strongly Agree were aligned to the same 0-4 structure. This ensured that all survey items carried a comparable numerical weight, where higher values consistently represented stronger agreement or greater frequency of engagement. Secondly, after encoding, numerical scores were aggregated at the variable level to compute average values. This approach allowed the representation of central tendencies across respondents while preserving ordinal meaning. **Table 1** shows a category reference describing the encoding scheme and its associated meaning for each survey construct.

*Table 1: Categorical scale for Encoded Survey Responses and the associated interpretations.*

| Aggregated values (0 - 4) | Description | Interpretation of Response Level |
|---|---|---|
| 0 - 0.49 | Strongly Disagree / Never | Complete rejection of the statement or behaviour; reflects the absence of the practice, perception, or experience |
| 0.5 - 1.49 | Disagree / Occasionally | Low acceptance or occurrence; the issue/practice is rare or viewed negatively. |
| 1.5 - 2.49 | Neutral / Regularly | Ambivalence or moderate occurrence; respondents neither lean strongly positive nor negative. |
| 2.5 - 3.49 | Agree / Frequently | Positive alignment: the practice, perception, or experience is common and generally supported. |
| 3.5 - 4 | Strongly Agree / Always | Very high endorsement; consistent presence of the practice, perception, or experience. |

Descriptive statistics were then used to summarise the overall response distribution. Frequency counts and percentages were computed for categorical and ordinal items to describe patterns in participant demographics and perceptions. These summaries provided a clear overview of the proportion of respondents in each category, highlighting dominant views and variations across survey dimensions.

## 3. Results and Discussion
### 3.1. Demographics and regional distribution of the survey

The survey recorded a total of 156 responses across participating countries. The country-level distribution of respondents indicates a strong concentration in Ghana, which accounted for 57.8% of all responses.

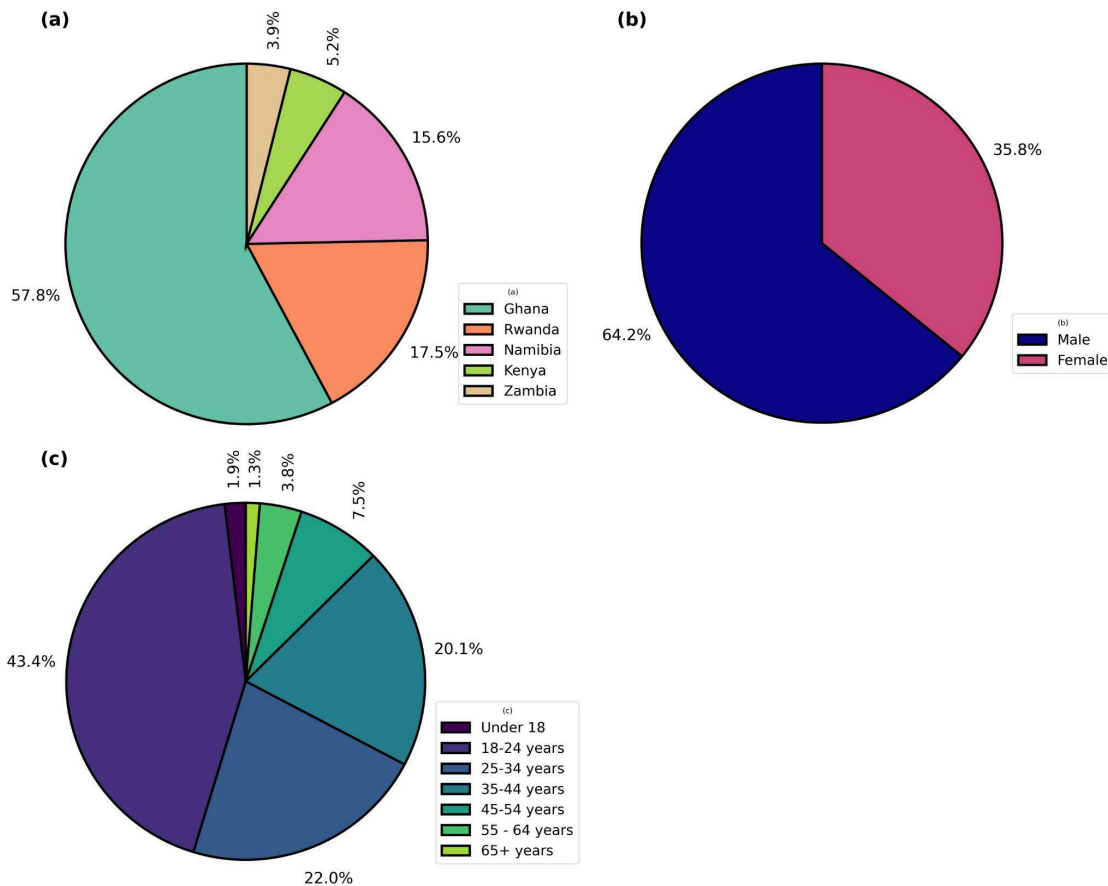

*Figure 1*: *Demographic distribution of survey respondents across country(a), gender identity (b), and age group (c).*

This was followed by Rwanda (17.5%) and Namibia (15.6%), while Kenya (5.2%) and Zambia (3.9%) represented smaller shares **(see Figure 1a)**. This imbalance is likely the result of the survey's wider circulation and greater visibility in Ghana, supported by local institutions and networks, which have improved accessibility and awareness compared to other countries. While the high concentration of responses from Ghana enhances the reliability of findings for that context, it constrains the comparative validity of insights from other countries. The gender distribution shows that male respondents constituted the majority at 64.2%, compared to 35.8% for female respondents, indicating higher male participation in the survey. This disparity may reflect broader gender gaps in digital access and participation, as well as the influence of the professional and institutional networks through which the survey was disseminated. Similar results were also evident in **Russo et al., 2025**, where their study highlighted that women showed higher AI anxiety, lower positive attitudes toward AI, lower use of AI, and lower perceived knowledge of AI. The lower participation of women highlights the need to consider gender inclusivity in future data collection efforts to ensure balanced representation.

The age distribution of respondents further shows a strong skew toward younger participants. The 18-24 years group accounted for 69 responses, forming the largest segment of the sample. The 25-34 years (35 responses) and 35-44 years (32 responses) groups also provided notable input, while respondents aged 45 years and above contributed relatively small proportions. Only three participants were under 18, and two were above 65 years old. The higher engagement of younger respondents in this survey is consistent with other African studies showing that greater digital literacy, exposure to online platforms, and interest in academic and professional themes drive stronger participation (**Ayisi et al., 2024**; **Afrobarometer Report, 2020**).

The findings underscore three salient patterns. First, the predominance of responses from Ghana reflects a pronounced geographic concentration within the sample. Second, the gender distribution reveals a marked underrepresentation of women. Third, the age distribution illustrates the strong participation of younger cohorts. Collectively, these imbalances provide critical context for interpreting the survey outcomes and underscore the need for future research designs to enhance inclusivity and achieve greater demographic and geographic representativeness.

### 3.2. Curriculum alignment and workforce readiness

The extent to which AI topics are incorporated into university curricula reflects both institutional capacity and responsiveness to evolving industry demand. Figure 2 illustrates the distribution of AI focus areas across five African countries based on survey responses expressed as percentages.

In Ghana, Machine Learning is the predominant area, cited in 38.7% of responses, followed by Natural Language Processing (15.3%), Artificial Neural Networks (13.5%), and Robotics (11.7%). Kenya demonstrates an even stronger concentration, with Machine Learning alone accounting for half of all responses (50%) **(Owino, 2023)**. Namibia displays a more balanced profile, where Machine Learning represents 32.4% and Robotics receives notable emphasis at 21.6%. Rwanda shows a diverse spread, with Machine Learning (28.3%), Artificial Neural Networks (23.9%), and Computer Vision (17.4%) all receiving attention. By contrast, Zambia stands out with a distinct emphasis on Computer Vision, which dominates at 66.7%, while Machine Learning constitutes the remaining 33.3%. This points out distinct country-level priorities in AI research and application. Ghana and Kenya's strong focus on Machine Learning indicates an emphasis on foundational AI techniques that cut across multiple application areas. Namibia's relatively higher share of Robotics points to interest in automation and applied AI solutions. Rwanda's spread across Neural Networks, Machine Learning, and Computer Vision suggests a diversified approach to AI capacity development. In contrast, Zambia's concentration on Computer Vision highlights a specific niche, possibly reflecting sectoral demands such as agriculture, surveillance, or medical imaging.

The perceived relevance of AI-related courses provides insight into how well institutional training aligns with industry needs. Respondents expressed varying levels of confidence in this alignment. As presented in Figure 3, the cross-tabulation between course relevance and job readiness reveals important patterns. Students who rated their AI courses as very relevant to industry demands showed a more even distribution in their perception of preparedness, with 41% reporting neutral confidence, 28% agreeing they are prepared, and 22% disagreeing.

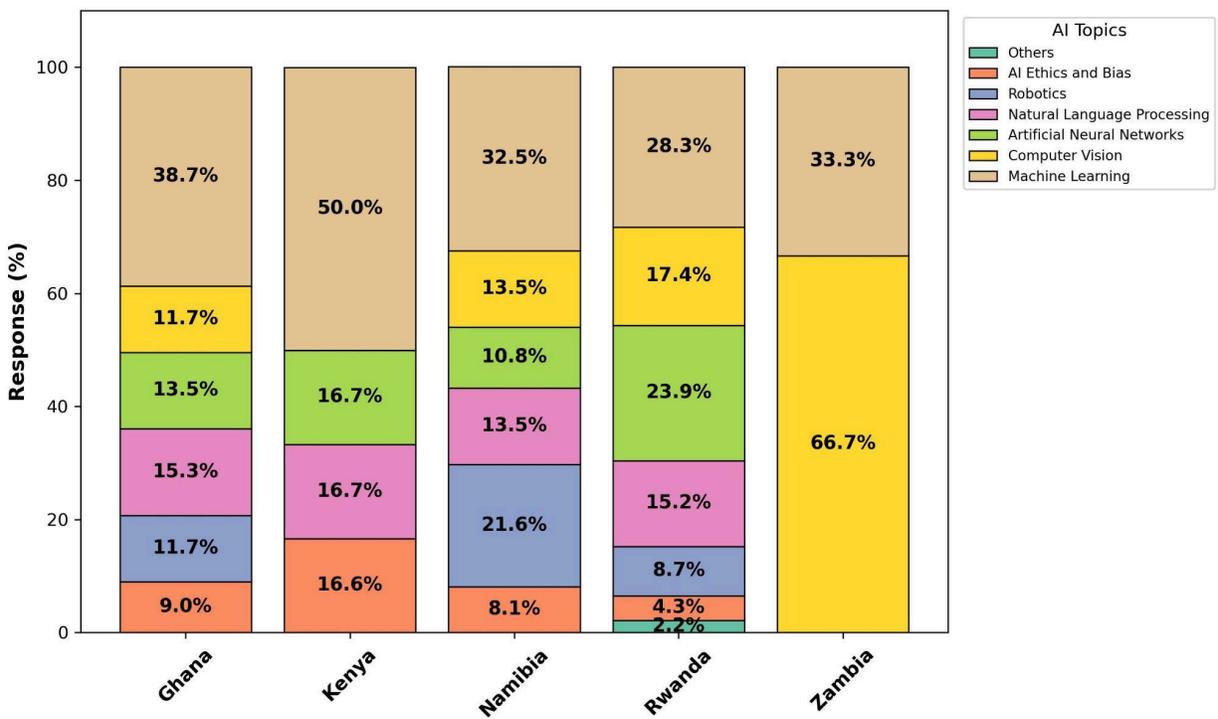

*Figure 2*: *Distribution of survey responses on AI topic focus across Ghana, Kenya, Namibia, Rwanda, and Zambia.*

This suggests that even when courses are perceived as aligned, not all students feel sufficiently prepared for the job market. Among those who rated their courses as somewhat relevant, 40% still felt neutral, while 23% agreed, and another 23% disagreed. This indicates a divided perception, reflecting moderate confidence but also highlighting gaps in translating course content into perceived job readiness. Students who described courses as neutral in relevance were the least confident, with the largest share (39%) maintaining neutral readiness and 34% expressing disagreement with being job-ready. Similarly, those who rated courses as somewhat irrelevant unanimously reported neutral confidence, indicating a lack of strong conviction in their skills.

Overall, the results demonstrate a clear link between perceived curriculum relevance and job market confidence. Students who view their courses as more relevant are more likely to feel prepared, though uncertainty remains widespread. The prevalence of neutral responses across all groups points to underlying challenges in connecting theory with practice, reinforcing the need for stronger experiential learning, applied projects, and institutional support to build job-ready AI skills. For policymakers and educators, such perceptions highlight the need to strengthen linkages between higher education curricula and the rapidly evolving demands of the AI workforce in Africa.

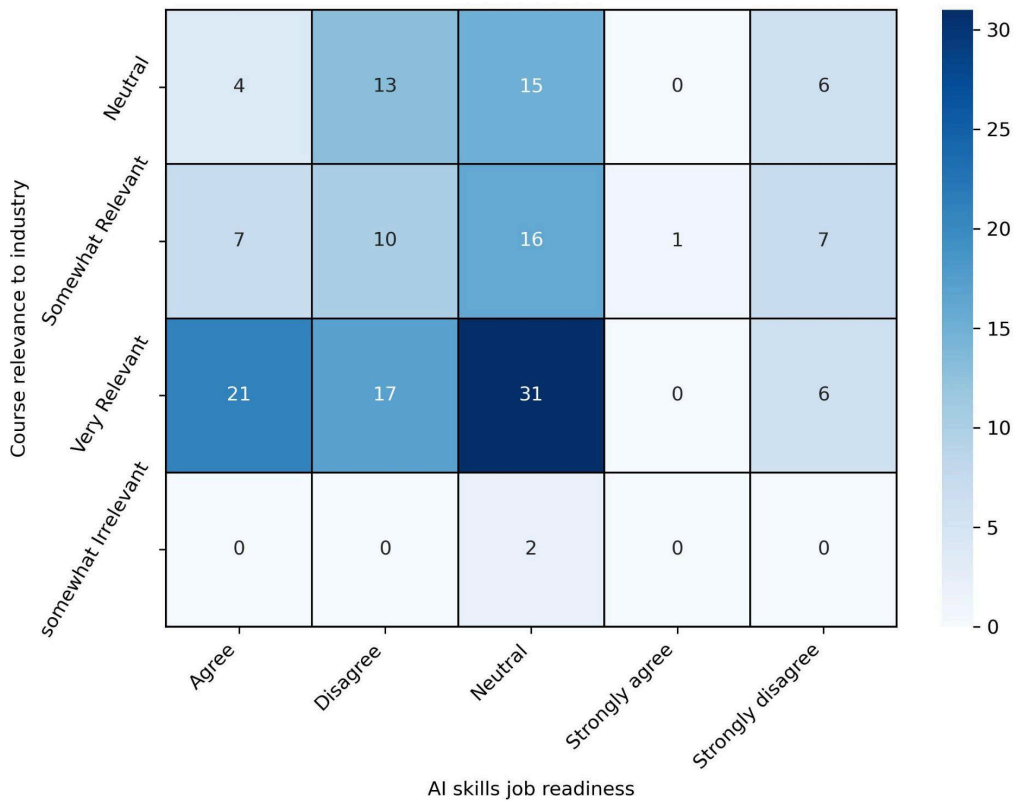

*Figure 3*: Relationship between students' perception of AI course relevance to industry and their self-assessed job market readiness.

### 3.3. Perceptions, Engagement and Resources in University-Industry AI Education

University-industry linkages are critical for strengthening AI education, as they shape how learners and staff perceive the relevance of AI, engage in practical activities, and access the resources needed for skill development. Figure 4 highlights the responses from participants, summarising their belief in AI's impact, engagement in research or coursework, participation in AI activities, and access to AI learning resources. General respondents across all employment categories (students, contract, full and part-time workers) and country expressed strong agreement that AI will shape their field in the next 5-10 years (see Figure 4a). Contract or temporary staff reported the highest average, above 3, which falls in the "Agree" range. Full-time staff and students also scored above 2.5, while part-time staff were slightly lower but remained positive. Awareness of AI's strategic importance is therefore well established in both academic and professional groups. Awareness of AI as a strategic asset is widely recognised in both academic research and professional practice globally, including in Africa, where countries are increasingly acknowledging AI's transformative potential for economic development, governance, and service delivery (**Csaszar et al., 2024**; **Borges et al., 2021**).

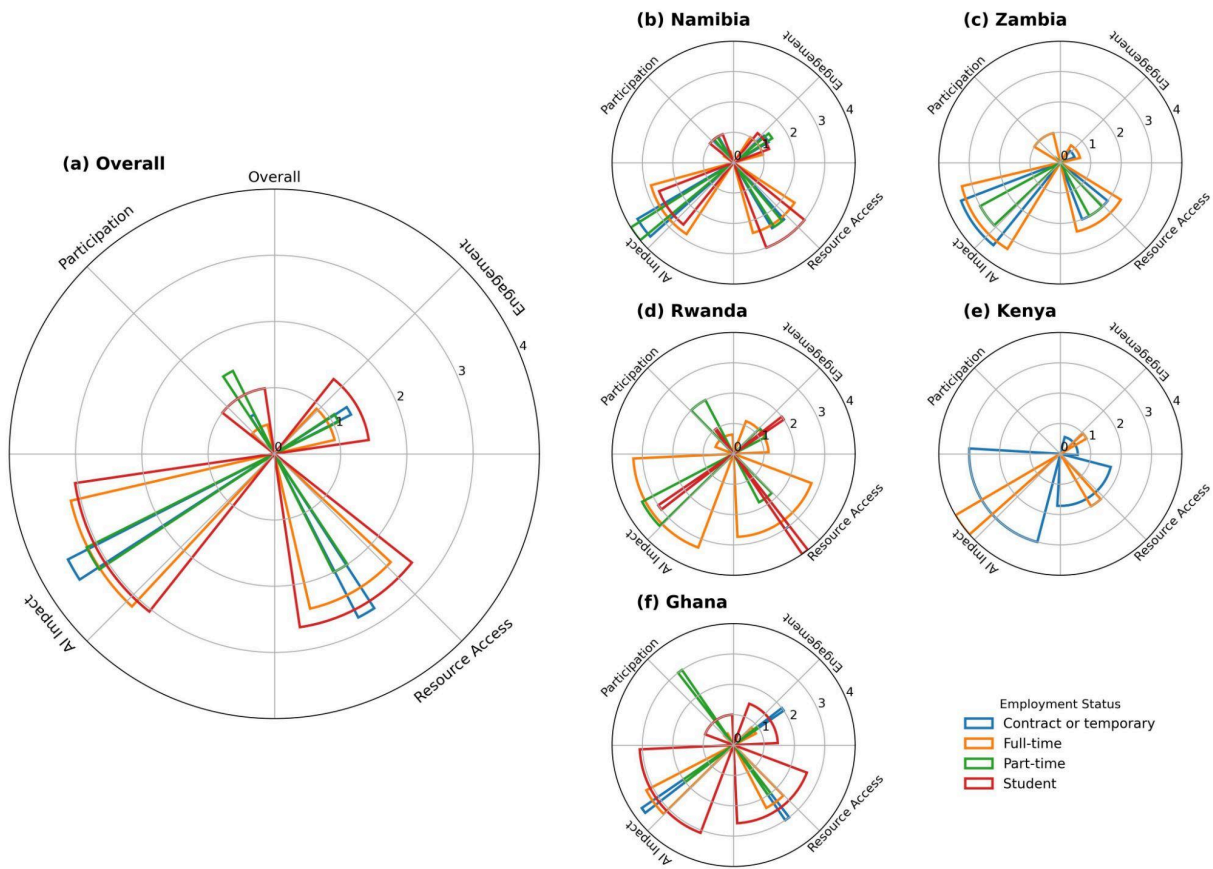

*Figure 4: Perceptions and practices of AI education and workforce preparedness across employment categories (Contract or temporary, Full-time, Part-time, Student). Values are displayed on a 0-4 scale, indicating levels of engagement, participation, AI tool use, perceived AI impact, and access to resources.*

AI is seen as a crucial enabler in strategic decision-making, enhancing organisational performance by augmenting cognitive capacities for analysis, prediction, and innovation (**Csaszar et al., 2024**). This strategic importance is reflected in investment trends and the formulation of national AI strategies across African countries, which shows that public and private sectors are integrating AI awareness into policy and practice (**Azaroual, 2024**). The country-level results show both alignment and some departures from this general pattern, as shown in Figure 4b - f. Overall, most countries show higher agreement among contract staff, but Namibia and Kenya stand out, with part-time staff reporting the strongest belief in AI's impact.

Despite high expectations for AI's impact, the general engagement and participation levels were significantly lower across various employment statuses and countries. Students averaged 1.43 for engagement in AI-related coursework or research, and only slightly higher for participation (0.98) in activities. This implies that they are occasionally engaged with or participated in AI-related projects or coursework. Part-time staff reported moderate involvement, close to neutrality, while contract and full-time staff achieved better results but stayed below 2.5. These findings reveal a clear gap between recognising AI's relevance and consistently taking part in related tasks or studies.

The response from different countries shows that the low overall engagement and participation levels were not uniform. Ghana, for instance, showed the clearest contrast: students had the highest engagement (1.45) but only participated in AI-related events occasionally (1.0), while part-time staff recorded frequent participation (3.0) despite their occasional engagement (1.0) as indicated in Figure 4b. Kenya had no response from students and contract workers, but part-time staff reported higher engagement (1.0) than full-time staff (0.57). Namibia stood out for stronger scores: part-time staff led in engagement (1.5) and participation (1.0), contract staff were close, while students were more active in participation (1.0) than full-time staff (0.38).

Access to resources showed the largest differences among groups. Contract and full-time staff reported averages near or above 3, indicating frequent availability of books, online courses, or computing facilities. Students scored around 2.5, suggesting that while materials exist, their availability or affordability is uneven. Part-time staff were slightly lower, reflecting limited institutional support or competing commitments. Resource constraints appear most pronounced for students and part-time personnel, potentially limiting their ability to apply knowledge in practice.

The overall pattern shows that while respondents perceive AI as highly relevant, active engagement, participation, and equitable access lag behind. Closer collaboration between universities and industry offers a way to bridge this divide. Internships, applied research placements, and jointly designed curricula can provide students and staff with authentic AI experience. Partnerships that give shared access to data, computing capacity, and proprietary tools would further reduce barriers. Strengthening these linkages is critical for transforming awareness of AI into meaningful participation and workforce-ready expertise.

## 3.4. Barriers and Enablers of Equitable AI Education

Understanding the barriers that limit access to AI education is essential for developing effective university-industry partnerships in Africa. Survey responses reveal several structural, financial, and informational challenges that prevent students and professionals from fully participating in AI training and related opportunities.

Analysis of the 156 responses shows that limited financial resources are the most reported obstacle. About half of the respondents noted that tuition fees, certification costs, or the price of supporting materials make it difficult to join AI courses. This finding aligns with documented financial constraints faced by African universities in providing accessible AI education (**Akinwalere & Ivanov, 2022**; **Azaroual, 2024**). Additional hidden costs, such as internet data charges, transport, and access to computing resources, further restrict participation, reinforcing the multifaceted nature of financial barriers (**Azaroual, 2024**). Over one-third of respondents cited unreliable or expensive internet as a significant barrier to engaging with AI content or enrolling in online courses, a challenge consistent with the digital divide documented in African educational contexts **(Borges et al., 2021)**. Comments about inadequate access to stable electricity or modern devices further support the infrastructural challenges identified in prior studies (**Azaroual, 2024**).

Additionally, nearly half of the participants reported being unaware of available programs, discussions, or forums. This gap affects students and professionals alike, suggesting that communication between universities, training providers, and industry remains fragmented or operating in silos. Without clear pathways and visibility of opportunities, potential students remain disconnected from AI education. Time constraints and high event fees also discourage participation in AI workshops and conferences, hindering exposure to practical AI applications (**Azaroual, 2024**). Many respondents reported that work or academic schedules prevent them from attending AI activities such as workshops, hackathons, or conferences. High event fees compound the issue, discouraging participation in discussions or forums that could broaden exposure to practical AI applications. Although only a few respondents reported explicit exclusion or discrimination, concerns remain regarding accessibility for rural, low-income, or non-technical populations. These social and structural factors limiting confidence and uptake reflect known barriers to equitable AI education in Africa.

Despite these constraints, respondents identified clear measures to strengthen participation. Inclusion in AI-related education varied, with only 18% feeling fully included and 23% excluded. Targeted initiatives for underrepresented groups were inconsistent, reported by just 22%, while 32% said none were in place. Learning resources were often rated as only moderately available, with 35% giving neutral assessments and just 16% describing them as highly accessible. When asked what would make AI education more effective, respondents highlighted industry partnerships and internships (76%), better computing and laboratory facilities (69%), scholarships or financial aid (68%), and hands-on projects (67%). Mentorship (63%) and inclusive teaching materials (60%) were also valued. These recommendations

13resonate with best practices advocated in recent research to bridge AI education gaps in Africa (**Yang & Matos, 2025**).

### 3.5. Policy

The survey results reveal varied levels of awareness, preparedness, and concern about AI policy and governance among respondents across African institutions and countries. Survey results indicate limited familiarity with AI-related policies and regulations among respondents. A majority (49% respondents) reported not being familiar with such frameworks, while 24% indicated moderate familiarity and 23% reported being somewhat familiar. Only 4% of participants described themselves as very familiar or expert. This distribution suggests a generally low level of awareness and engagement with AI policy across the sample population. When asked whether their country has a national AI policy or strategy, 69 % of the respondents reported being unsure, 17% indicated that their country does not have one, and only 14% confirmed the existence of such a framework. The high proportion of uncertain respondents highlights a significant gap in communication and dissemination of national AI initiatives. It further suggests that even where national AI strategies or policies exist, awareness among practitioners, students, and institutional stakeholders remains limited. Overall, these patterns are consistent with recent studies describing Africa's AI governance landscape as emergent and uneven. Existing evidence shows that while several African countries are advancing national AI strategies, only a limited number have formalised comprehensive AI policies. As of 2025, at least twelve countries have adopted national AI strategies, but only three have implemented fully developed AI policy frameworks (**Munga & Quansah, 2025**). This trend reflects a transitional governance stage in which strategic frameworks are serving as precursors to binding policy and regulatory instruments aimed at ensuring responsible and inclusive AI development (**Kwarkye, 2025**; **Munga & Quansah, 2025**).

Within this evolving governance environment, issues of inclusiveness remain a notable concern. Responses revealed moderate confidence in the extent to which AI policies in institutions or countries consider the needs of vulnerable and marginalised groups. Only 10% of the respondents expressed high confidence in the inclusivity of AI policies in their country or workplace, reflecting the known challenges in inclusivity within AI governance frameworks in Africa (**Plantinga, 2024**; **Azaroual, 2024**). These findings suggest that inclusivity and equity considerations are not yet perceived as central features of AI policy implementation within the African context, a pattern that is also evident in AI4D policy documents, which emphasise strategic objectives but provide limited actionable frameworks for addressing equity and representation.

The survey results revealed deeper insights into perceptions of ethics, regulation, and governance in AI adoption across Africa. Respondents frequently cited data privacy, bias and discrimination, misinformation, lack of transparency, and job displacement as key ethical concerns. Data privacy and misinformation dominated responses, reflecting growing public sensitivity to data misuse and information integrity. These concerns mirror regional analyses that highlight weak data protection systems and limited enforcement capacity (**Kaddu & Ssekitto, 2023**). Perceptions of regulation were mixed. Nearly half of the respondents were unsure about the adequacy of AI regulations in their countries, and only a few respondents considered them adequate. This uncertainty indicates limited policy visibility and weak institutional capacity for AI governance. Similarly, most respondents reported that their institutions had not adopted AI ethics guidelines or responsible AI frameworks, reinforcing the gap between policy discourse and practice.

When asked about responsibility for AI governance, most respondents emphasised governments, followed by academic institutions and regional bodies such as the African Union and ECOWAS. This reflects expectations for state-led coordination but also highlights the need for stronger multi-stakeholder engagement. Confidence in Africa's ability to develop its own AI regulatory frameworks was moderate, suggesting recognition of regional initiatives but persistent concerns about resource and implementation challenges. The African Union's Continental Artificial Intelligence Strategy, adopted in 2024, sets a framework for responsible AI development and deployment across the region (**African Union, 2024**). Complementary initiatives such as Smart Africa and the partnership on AI promote cross-border collaboration and knowledge sharing to ensure AI governance aligns with Africa's socio-economic priorities (**Plantinga, 2024**). This was highlighted in the survey, where participants showed broad agreement on the importance of tailoring AI regulation to Africa's socio-economic and cultural contexts, and on Africa playing an active role as a leader or contributor in global AI governance discussions. Respondents further emphasised the need for concrete governance actions, prioritising stronger data protection laws, ethical AI certification, fairness audits, and public engagement programs. Together, these perspectives underscore growing awareness of the importance of developing context-sensitive and participatory regulatory frameworks that balance innovation with social accountability across the continent.

4. **Conclusion**

This study presents an evidence-based perspective on how African students and professionals experience AI education, the resources and opportunities available to them, and the barriers and enablers that shape equitable access to this education. Across five countries, respondents recognised AI's strategic importance but reported limited engagement with structured training, low participation in activities, and uneven access to resources. Curriculum relevance to workforce demands improved perceived job readiness, yet persistent gaps between theory and practice highlight the need for applied learning and stronger university-industry collaboration. Financial constraints, weak infrastructure, and limited awareness emerged as major obstacles to participation, which were further compounded by fragmented communication between institutions and training providers. Resource challenges, including unreliable internet and insufficient computing facilities, disproportionately affected students and part-time staff.

Explicit exclusion was rare, but perceptions of inaccessibility among rural, low-income, or non-technical groups suggest the existence of deeper structural inequities.

Respondents identified several clear enablers, including targeted support for underrepresented groups, improved access to learning materials, internships, industry-linked projects, and mentorship. Effective university-industry partnerships, backed by scholarships and inclusive curricula, were viewed as essential for building practical skills and broadening participation.

Policy awareness and confidence in governance varied widely, with respondents emphasising transparency, accountability, and capacity development as priorities for sustainable AI adoption. National strategies and institutional ethics guidelines remain unevenly developed, and attention to vulnerable groups is often lacking. Strengthening AI education in Africa will require closing the gap between awareness and participation, aligning curricula with labour market needs, and addressing financial and infrastructural barriers. Effective AI ecosystem building in Africa requires simultaneous attention to three dimensions: policy coordination, institutional capacity, and ethical governance. Coordinated action among governments, universities, and private industry will determine whether the continent moves from fragmented initiatives toward a sustainable, context-aware AI education and innovation framework capable of serving its long-term socio-economic goals.

**Acknowledgement**

This study has been ably funded by the Microsoft AI Economy Institute, an initiative of the Microsoft AI for Good Lab. The authors sincerely appreciate the invaluable contributions of all **stakeholders, collaborators, and partner institutions** across **Rwanda, Namibia, and Ghana**, whose engagement, insights, and shared commitment greatly enriched the design and implementation of this work.